\documentclass[aps,prl,twocolumn,superscriptaddress]{revtex4}

\usepackage[dvips]{graphicx}
\usepackage{color}
\usepackage{bm}
\usepackage{amssymb}
\renewcommand{\vec}[1]{\boldsymbol{#1}}
\usepackage[utf8]{inputenc}
\usepackage[version=4]{mhchem}
\usepackage{blindtext}
\usepackage{array}
\usepackage{multirow}

\usepackage{hyperref}

\newcommand{\etal}{\textit{et al}.}

\begin{document}

\title{Canted Antiferromagnetic Order in the Kagome Material Sr-Vesignieite} 

\author{A. Verrier}
\affiliation{Institut Quantique and D\'{e}partement de Physique, Universit\'{e} de Sherbrooke, 2500 boulevard de l'Universit\'{e}, Sherbrooke, Qu\'{e}bec J1K 2R1, Canada}
\author{F. Bert}
\affiliation{Laboratoire de Physique des Solides, CNRS, Univ. Paris-Sud, Universit\'{e} Paris-Saclay, 91405 Orsay Cedex, France}
\author{J.-M. Parent}
\author{M. El-Amine}
\affiliation{Institut Quantique and D\'{e}partement de Physique, Universit\'{e} de Sherbrooke, 2500 boulevard de l'Universit\'{e}, Sherbrooke, Qu\'{e}bec J1K 2R1, Canada}
\author{J. C. Orain}
\affiliation{Laboratoire de Physique des Solides, CNRS, Univ. Paris-Sud, Universit\'{e} Paris-Saclay, 91405 Orsay Cedex, France}
\author{D. Boldrin}
\author{A. S. Wills}
\affiliation{Department of Chemistry, University College London, 20 Gordon Street, London, WC1H 0AJ, United Kingdom}
\author{P. Mendels}
\affiliation{Laboratoire de Physique des Solides, CNRS, Univ. Paris-Sud, Universit\'{e} Paris-Saclay, 91405 Orsay Cedex, France}
\author{J. A. Quilliam}
\email[]{Jeffrey.Quilliam@USherbrooke.ca}
\affiliation{Institut Quantique and D\'{e}partement de Physique, Universit\'{e} de Sherbrooke, 2500 boulevard de l'Universit\'{e}, Sherbrooke, Qu\'{e}bec J1K 2R1, Canada}

\date{\today}

\begin{abstract}
We report $^{51}\mathrm{V}$ NMR, $\mu \mathrm{SR}$ and zero applied field $^{63,65}\mathrm{Cu}$ NMR measurements on powder samples of Sr-vesignieite, $\mathrm{SrCu}_3\mathrm{V}_2\mathrm{O}_8\mathrm{(OH)}_2$, a $S=1/2$ nearly-kagome Heisenberg antiferromagnet. Our results demonstrate that the ground state is a $\vec{q}=\vec{0}$ magnetic structure with spins canting either in or out of the kagome plane, giving rise to weak ferromagnetism. We determine the size of ordered moments and the angle of canting for different possible $\vec{q}=\vec{0}$ structures and orbital scenarios, thereby providing insight into the role of the Dzyaloshinskii-Moriya (DM) interaction in this material.
\end{abstract}


\maketitle

\section{INTRODUCTION}

It is now fairly well accepted from a theoretical point of view that the ideal spin-1/2, nearest-neighbor, Heisenberg kagome antiferromagnet (KAFM) should have a quantum spin liquid (QSL) ground state~\cite{Yan2011}. There is also very encouraging evidence that this does not only occur in the perfect theoretical toy model, but is also borne out in real material systems. Notably, ZnCu$_3$(OH)$_6$Cl$_2$ (herbertsmithite)~\cite{Mendels2007,Olariu2008,Fu2015,Han2012Nature} as well as several variants such as ZnCu$_3$(OH)$_6$FBr~\cite{Feng2017} and ZnCu$_3$(OH)$_6$SO$_4$~\cite{Gomilsek2016,Gomilsek2017}, are strong candidates for kagome QSL ground states. 

An important consideration for realistic kagome compounds, however, is the Dzyalloshinskii-Moriya (DM) interaction, $\vec{D}_{ij}\cdot\left(\vec{S}_i\times\vec{S}_j\right)$, which is allowed by the symmetry of the kagome lattice. This interaction, if sufficiently strong, is understood to induce magnetic order~\cite{Cepas2008,Lee2018}. Even for QSL materials like herbertsmithite, the DM interaction is present~\cite{Zorko2008} and may have important consequences~\cite{Jeong2011,Messio2017}. The role of both non-zero components of the DM vector ($D_z$ out-of-plane and $D_p$ in-plane) and their effects on the chirality and weak ferromagnetism of the ordered ground state has not been thoroughly investigated from an experimental perspective due to a small number of appropriate materials.

\begin{figure*}[htb]
	\centering
	\begin{tabular}{ccc}
	\includegraphics[height=0.2\linewidth]{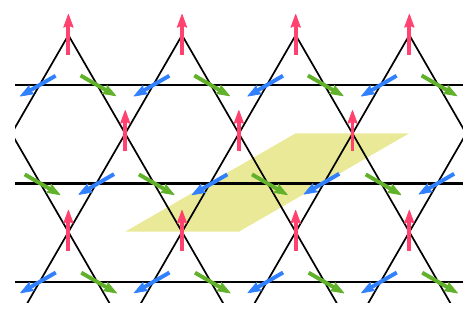} &\quad\quad &
	\includegraphics[height=0.2\linewidth]{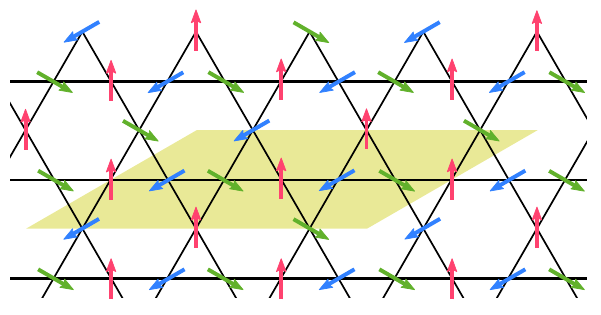} \\
	PVC & & SVC \\
	\\
	\includegraphics[height=0.2\linewidth]{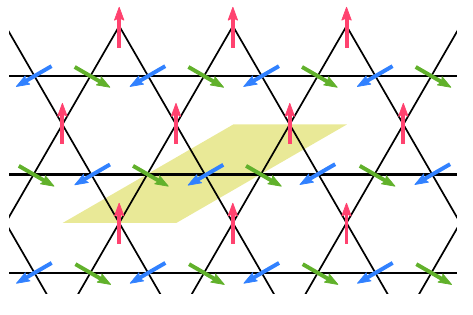} & \quad\quad&
	\includegraphics[height=0.2\linewidth]{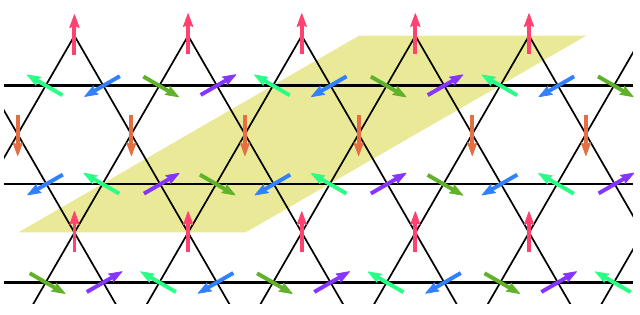} \\
	NVC & &Hexagonal triple-$\vec{q}$ \\
	\end{tabular}
	\caption{Different coplanar magnetic orders that have been proposed for kagome antiferromagnets and their respective unit cells. PVC, NVC and SVC are 120$^\circ$ phases and stand for positive, negative and staggered vector chirality, respectively, where the vector chirality is defined for every triangle as $\vec{\kappa}=\frac{2}{3}\left( \vec{S_1}\times\vec{S_2}+\vec{S_2}\times\vec{S_3}+\vec{S_3}\times\vec{S_1} \right)$ with labels increasing when turning counterclockwise around the triangle. The SVC order is also referred to as $\sqrt{3}\times\sqrt{3}$ or $\vec{q}=(1/3~ 1/3)$ in other works. The hexagonal triple-$\vec{q}$ state is described in more detail in Ref.~\cite{Boldrin2018}.}
	\label{fig:magnetic_orders}
\end{figure*}

One important test-case is the mineral Cu$_3$BaV$_2$O$_8$(OH)$_2$, or vesignieite, which has $S=1/2$ Cu$^{2+}$ spins on an almost perfect kagome lattice, with only a 0.2\%~\cite{Okamoto2009} to 0.5\%~\cite{Boldrin2016} bond-length asymmetry. While this material was initially proposed as a QSL candidate, later measurements showed a coexistence of weak magnetic order and persistent spin fluctuations~\cite{Colman2011,Quilliam2011Vesig}, suggesting that it had a DM interaction just past the critical point between spin liquid and antiferromagnetism. In samples with higher crystallinity, it has been found that the ordered magnetic moment is not particularly small~\cite{Yoshida2012} implying that the DM interaction may also be quite large. On the basis of NMR measurements, Yoshida \etal ~claimed that the material exhibited $\vec{q}=\vec{0}$ magnetic order with positive vector chirality (PVC) as shown in Fig.~\ref{fig:magnetic_orders} with slight out-of-plane canting.

On a vesignieite pseudocrystal, that order was subsequently proposed to have negative vector chirality (NVC, see Fig.~\ref{fig:magnetic_orders}) with slight canting in the kagome plane~\cite{Ishikawa2017}, similar to what was observed in another kagome compound Cd-kapellasite~\cite{Okuma2017}. Moreover, ESR measurements have shown that while $D_z$ is of similar magnitude in herbersmithite and vesignieite, $D_p$ is much more significant in vesignieite and is likely mainly responsible for the ordering~\cite{Zorko2013}. However, this last conclusion was made under the assumption that the main interactions were dominantly nearest-neighbor. In parallel, the $P3_1 2 1$ space group was proposed~\cite{Boldrin2016} to provide a better fit to X-ray diffraction data than the previously accepted $C 2/m$ symmetry. The group working on a pseudocrystal of vesiginieite~\cite{Ishikawa2017} found no improvement in the refinement, hence, the crystalline structure of vesignieite remains contentious. Most recently, Boldrin \etal~\cite{Boldrin2018} have proposed a completely different, ``hexagonal triple-$\vec{q}$'' ground state for vesignieite, also shown in Fig.~\ref{fig:magnetic_orders}.

In this article, we study a kagome system related to vesignieite: Cu$_3$SrV$_2$O$_8$(OH)$_2$, or ``Sr-vesigineite'', which orders magnetically at low temperatures~\cite{Boldrin2015}. From this point forward, we refer to the original vesiginieite material as ``Ba-vesignieite'' in order to avoid confusion with the material studied here, ``Sr-vesignieite''. This study aims to provide further insight into the nature of the magnetic order in this material and the role of the DM interaction. Employing NMR and $\mu$SR measurements, we show that Sr-vesignieite orders in a spin configuration that is consistent with a $\vec{q}=\vec{0}$, 120$^\circ$ order and we rule out hypothetical orders such as SVC (staggered vector chirality, see Fig.~\ref{fig:magnetic_orders}). Based on the distinct internal fields observed at the $^{51}$V and $^{63,65}$Cu sites, we can estimate the size of the ordered moment and the level of canting of the spins. Most importantly, we find that given identical magnetic structures and orbitals, the ordered moment is smaller in Sr-vesignieite than in the original Ba-vesignieite, implying that it may provide a useful mid-point between herbertsmithite and Ba-vesignieite in the DM-induced phase diagram of the $S=1/2$ kagome antiferromagnet.

\section{EXPERIMENT}

Samples of Sr-vesignieite were prepared as described in Ref.~\cite{Boldrin2015}. Muon spin rotation ($\mu$SR) measurements were performed in zero-field and weak transverse-field geometries using the GPS instrument at PSI's $\mathrm{S\mu S}$ over a temperature range from 1.6 to 20 K. 

$^{63,65}\mathrm{Cu}$ NMR measurements were performed in zero applied field. The spectra were obtained by sweeping the resonance frequency of the typical RLC circuit for NMR together with the carrier frequency of pulses sent using a Tecmag \emph{Redstone} spectrometer. A $\pi/2-\tau-\pi$ (Hahn) sequence was used. Due to limitations of the adjustable capacitors, several RF coils were used to cover the entire frequency range studied. $^{51}\mathrm{V}$ NMR spectra were obtained by reconstruction at magnetic fields ranging from 2.7~T to 3.3~T, using a central frequency of 33.7~MHz. The reconstruction process consisted of four steps. The first step is to compute the Fourier transform of the echo measured at every applied field and to center it around the central frequency rather than 0. As a second step, the frequencies of these Fourier transforms are rescaled by the application $f\rightarrow f\frac{B}{B^\prime}$, where $B^\prime$ is the actual applied magnetic field and $B$ is the field at which the spectrum is being reconstructed. Summing these rescaled Fourier transforms is now a non-trivial matter, since the frequencies are now misaligned. To overcome this difficulty, the third step is to compute a linear interpolation function for every rescaled Fourier transform and these functions are summed. This reconstruction technique is similar to the one described in Ref.~\cite{Clark1995}.

Spin-lattice relaxation rate, $1/T_1$, measurements were performed using a $\pi-\mathcal{T}-\pi/2-\tau-\pi$ (inversion-recovery) sequence and by fitting the echo intensity with the following equation~\cite{exponential} where $M_0\sim 2M_{\mathrm{eq}}$:
\begin{equation}
M(\mathcal{T})=M_{\mathrm{eq}}-M_0\exp{\left[-\left(\frac{\mathcal{T}}{T_1}\right)^\beta\right]}.
\end{equation}

\begin{figure}
  \includegraphics[width=0.9\linewidth]{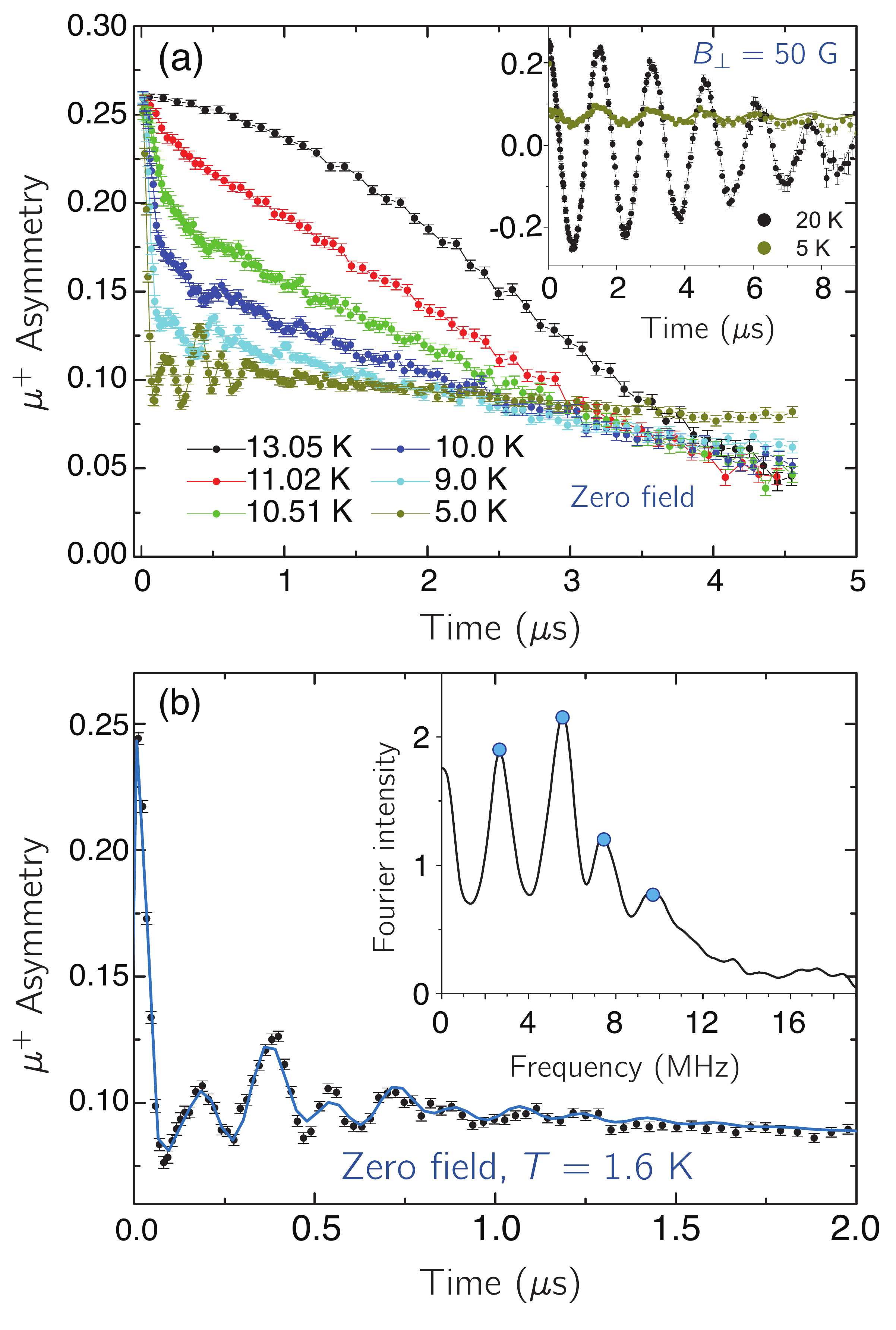}
  \caption{(a) Zero-field muon asymmetry in Sr-vesignieite. Silver sample holder background is estimated to be $\sim 0.02(1)$ from the remaining oscillating asymmetry in low transverse field at 5 K well below $T_\mathrm{N}\simeq 11~\mathrm{K}$ (see inset). (b) Zero-field asymmetry at 1.6 K. Four spontaneous oscillation frequencies are needed to fit the asymmetry in the time domain. They can be readily identified in the Fourier transform of the asymmetry decay (see inset).}
  \label{fig:muSR}
\end{figure}

\section{RESULTS AND DISCUSSION}

In this discussion we will first present $\mu$SR results, followed by zero-field Cu NMR results. $^{51}$V NMR results will then be presented and used to rule out certain types of magnetic order. We will focus on magnetic orders that have been proposed for Ba-vesignieite and on SVC. The reason we include SVC is because it is a 120$^\circ$ order where spin canting would create weak ferromagnetism and an FC-ZFC splitting in bulk susceptibility~\cite{Boldrin2015}. Other magnetic orders like octahedral, cuboc1 and cuboc2~\cite{Messio2011} have opposing spins in equal numbers on every crystallographic site. When that condition is met, a canting of one spin will always be compensated by the same canting angle (giving the opposite change in magnetization) on the opposing spin, given that the source of canting is either anisotropy or DM interaction. We limited our analysis to what we consider the most credible magnetic orderings for Sr-vesignieite, thus we cannot rule out all other possible magnetic structures.

Finally, drawing on the results of both Cu and V NMR experiments, and under the assumption of $\vec{q}=\vec{0}$ order, we will extract a range of possible ordered moment sizes and canting angles.

\subsection{Muon Spin Rotation}

The results of zero-field $\mu$SR measurements are shown in Fig.~\ref{fig:muSR}. Clear muon oscillations provide unambiguous evidence of long-range magnetic order at low temperatures, the onset of which is found between 11 and 13 K. A Fourier transform of the data (shown in the inset of Fig.~\ref{fig:muSR}b) reveals 4 distinct oscillation frequencies. This number of resonances should be compared with the five inequivalent crystallographic oxygen sites in the unit cell, as muons have a tendency to stop close to strongly electronegative atoms. It may be that one of the peaks (probably the most intense one) is produced by two sites with similar environments and unresolved frequencies. Since sites O11 and O13 are similar to each other and dissimilar to the remaining oxygen sites, they might well account for the unresolved muon positions. The temperature at which spontaneous oscillations appear in the $\mu^+$ asymmetry is consistent with the reported Néel temperature in Ref.~\cite{Boldrin2015}. More precisely, in magnetic susceptibility measurements, an abrupt upturn occurs at around 12.5 to 13 K whereas a FC-ZFC splitting appears at around 11 K. A comparison of the asymmetry at low transverse field in the ordered and disordered phases shows that fewer than 10\% of the muons are stopping in a non frozen volume at low temperature. This small fraction likely reflects a small background signal from muons not implanted within the sample, while the sample itself appears to be fully long range ordered below the magnetic transition.

\begin{figure}
  \includegraphics[width=0.9\linewidth]{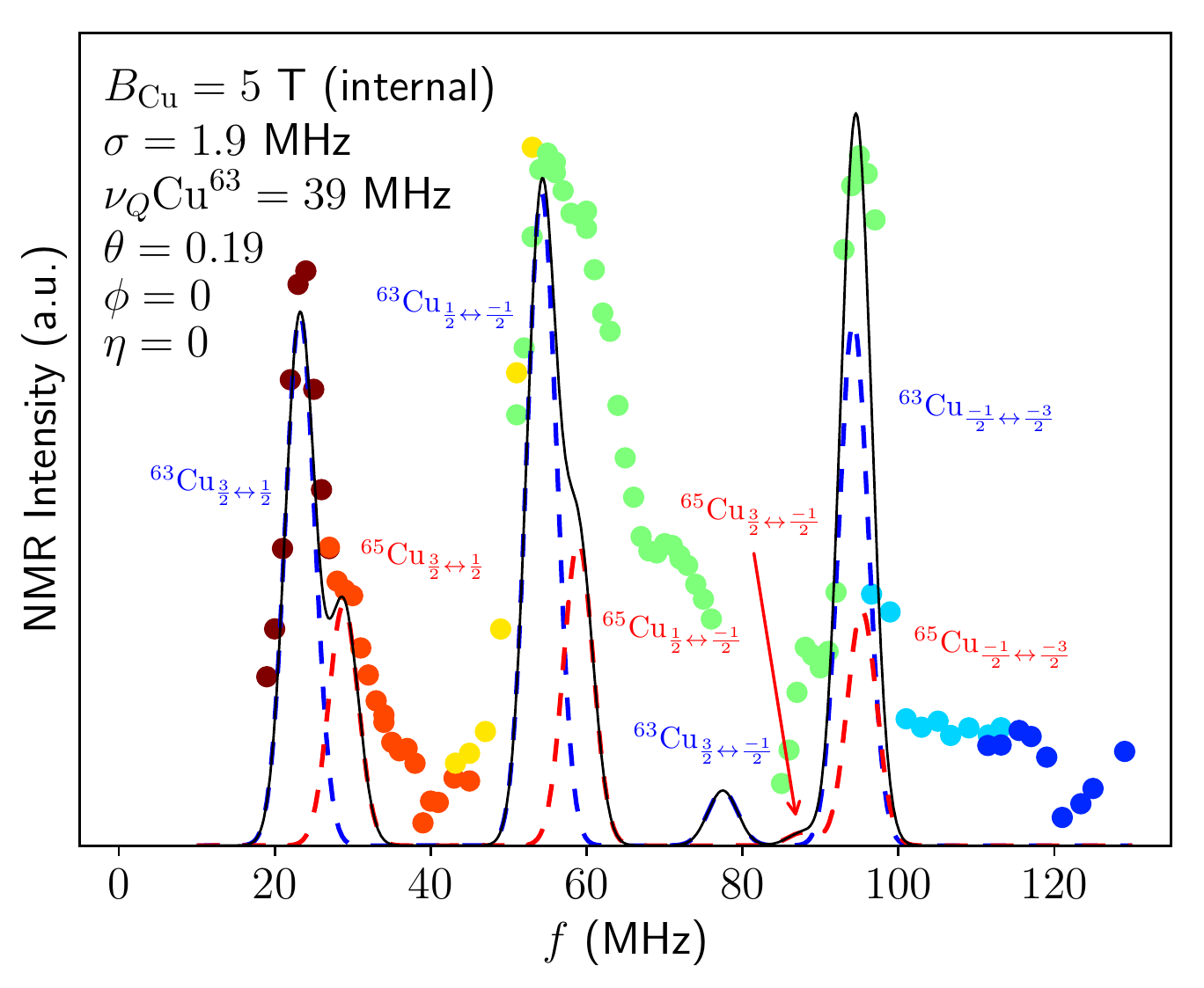}
  \caption{Zero-field NMR spectrum at 3.9 K. Each color corresponds to experiments performed with a particular RF coil. An example of simulation using exact diagonalization of Eq.~\ref{eq:ED} is shown, where the main peaks are in agreement with the data. The NMR intensity is weighted by a factor $f^{-1}Q^{-1}$ where $Q$ is the quality factor of the NMR RLC circuit.}
  \label{fig:CuSpectrum}
\end{figure}

\subsection{Zero-Field $^{63,65}$Cu NMR}

The zero-field NMR spectrum of Sr-vesignieite appears in Fig.~\ref{fig:CuSpectrum}. Since the magnetic moments are located on the Cu atoms, Cu nuclei are the ones encountering the most intense magnetic field, meaning that an external applied field is not necessary to obtain a significant NMR signal in the magnetically ordered phase. Like the zero-field $\mu$SR data, this spectrum provides direct evidence of spontaneous internal magnetic fields indicating long-range magnetic order. 

In considering the difference in frequency between the satellite peaks and the central peak of the spectrum, it becomes evident that neither the Zeeman energy nor the quadrupolar interaction is dominant. Thus, rather than using perturbation theory to model the spectrum, we used exact diagonalization of the following nuclear spin Hamiltonian:
\begin{align}
\mathcal{H}_Z=&-\gamma\hbar B_\mathrm{Cu}\left(\sin\theta\left( I_X\cos \phi +I_Y\sin\phi \right)+I_Z\cos \theta \right)\notag\\
\mathcal{H}_Q=&\pi\hbar\nu_Q\left( I_Z^2-\frac{1}{3}I^2+\frac{\eta}{3}\left( I_X^2-I_Y^2\right)\right)\notag\\
\mathcal{H}=&\mathcal{H}_Z+\mathcal{H}_Q
\label{eq:ED}
\end{align}
where $X$, $Y$ and $Z$ are the usual electric field gradient (EFG) coordinates. $I_X$, $I_Y$, $I_Z$ and $I^2$ are the $I=\frac{3}{2}$ spin operators. $B_\mathrm{int}$, $\theta$ and $\phi$ are fitting parameters corresponding to the spherical coordinates of the magnetic field at the Cu site relative to the principal axes of the EFG tensor. $\nu_Q^{Cu^{63}}$ and $\eta$ are additional fitting parameters describing the magnitude and asymmetry of the quadrupolar interaction. The eigenvalues of $\mathcal{H}$ were used as central frequencies of Lorentzian peaks (the width of which is left as a fitting parameter) and its eigenvectors were used to set the amplitude of these peaks by computing the matrix elements of a powder averaged Zeeman perturbative Hamiltonian in the eigenbasis. The ratio $\nu_Q^{\mathrm{Cu}^{63}}/\nu_Q^{\mathrm{Cu}^{65}}$ was kept constant at 1.082, based on the ratio of quadrupolar moments of the two isotopes. Due to the comparable importance of the quadrupolar and Zeeman energies, normally forbidden transitions can appear, such as the $\frac{3}{2}\rightarrow -\frac{1}{2}$ transition that is expected in this frequency range. 

The agreement between the main peaks of the experimental data and the model is satisfactory enough to deduce approximate values of 5~T and 39~MHz for $B_\mathrm{int}$ and $\nu_Q$ respectively. However, we failed to achieve good quantitative agreement and to explain some parts of the signal. In particular, there is significant spectral weight from 60 to 80 MHz and above 100 MHz that cannot be adequately explained by this simple model. This may be due to either insufficient exploration of parameter space or to an inaccurate model. Notably, one major assumption made is that the angle $\theta$ between the magnetic field and the unpaired orbital is taken to be the same for all Cu sites. If one considers PVC order and the same unpaired $d$-orbital is found on all Cu sites, this would be a valid assumption. However, Boldrin \emph{et al.}~\cite{Boldrin2015} propose that the Cu1 and Cu2 sites in Sr-vesignieite have rather different orbital physics. Based on observed Cu-O bond lengths, they propose that the Cu1 sites exhibit static $d_{x^2-y^2}$ orbitals, whereas the Cu2 sites exhibit a dynamic Jahn-Teller effect, with two energetically equivalent active $d_{x^2-y^2}$ orbitals. It may be that the missing spectral weight in our model could be accounted for by a very different spectrum arising on the Cu2 lattice site.

Alternatively, NVC order could also cause discrepancies with this simple model. Even in a perfect kagome lattice picture, with equivalent orbitals throughout, the NVC order only has a finite number of global spin rotations for which $\theta$ is fixed for all sites. In the general case, $\theta$ would vary from one site to another and this would yield more than three peaks per isotope. These changes might affect the spectrum through both the anisotropy of the hyperfine coupling and through the anisotropy of the quadrupolar interaction. Considering a more realistic model of PVC order with two inequivalent Cu sites or NVC order requires the introduction of a large number of free parameters. Although it might provide a highly over-parametrized fit, it would be unlikely to help us to distinguish between these two magnetic structures. The main accomplishment of the zero-field Cu spectrum obtained here is to provide a solid estimate of the internal field at the Cu-site, which allows us to place restrictions on the moment size and angle of spins.

\begin{figure*}
  \includegraphics[width=\linewidth]{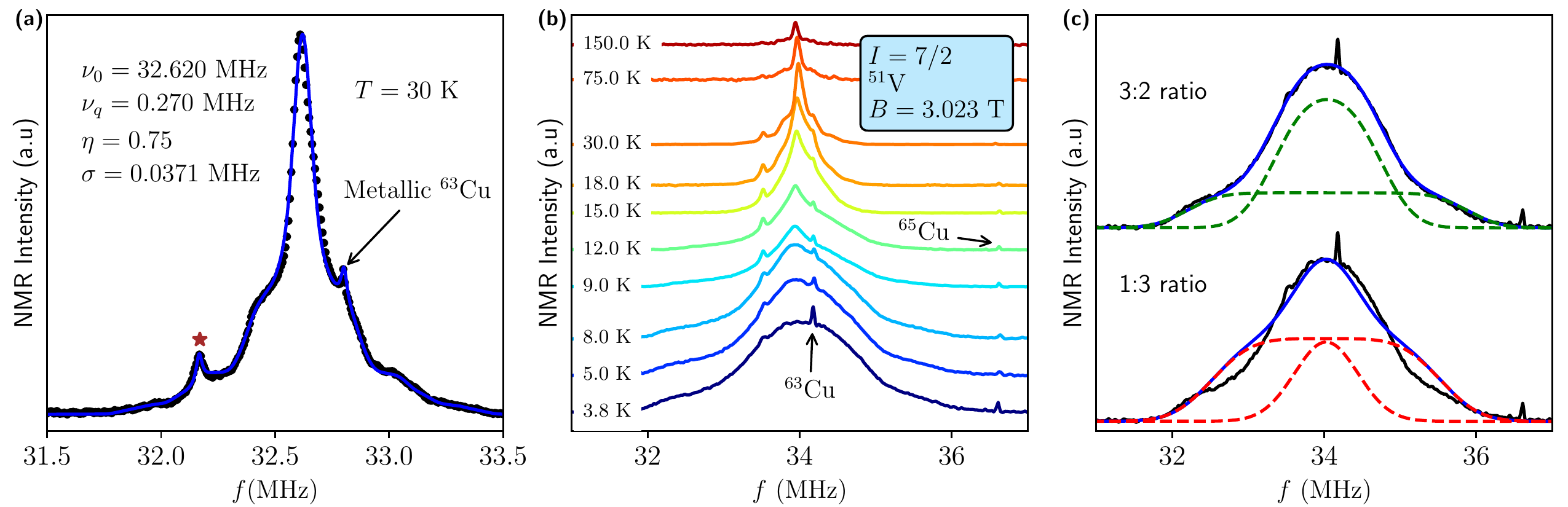}
  \caption{\textbf{(a)} Reconstruction of the $\ce{^{51}V}$ NMR spectrum at temperature well above the transition (points). A powder simulation of the quadrupolar perturbed Zeeman hamiltonian (solid line) gives excellent agreement with the data provided two narrow lorentzian peaks are added at 32.17~MHz and 32.8~MHz. The first additional peak presumably originates from non-magnetic impurities in the sample whereas the second is signal from copper in the rf coil. \textbf{(b)} Reconstructions of the $\ce{^{51}V}$ NMR spectra at varying temperatures. \textbf{(c)} Comparison of two simple models for the 3.8~K spectrum. For each model, two sites are assumed, giving rounded rectangular spectra of different widths (obtained through a convolution of a rectangular spectrum of varying width and a gaussian function of fixed width). Forcing the spectral weight of the narrow and broad sites to a 1:3 ratio, as would be consistent with the hexagonal triple-$\vec{q}$ order, yields a poor fit (bottom). In contrast, a good quality fit is obtained with a 3:2 ratio (top).}
  \label{fig:triplePlots}
\end{figure*}

\subsection{$^{51}$V NMR in applied field}

Fig.~\ref{fig:triplePlots} displays NMR spectrum reconstructions at different temperatures under applied fields of around 3~T. In the paramagnetic phase, a typical quadrupolar powder spectrum is observed. A fit to this data reveals a quadrupolar frequency of $\nu_Q = 0.270$ MHz and $\eta = 0.75$. A small, narrow peak marked with a star around 32.2 MHz is likely a non-magnetic V-containing impurity phase. A similar peak was also observed in Ba-vesignieite~\cite{Quilliam2011Vesig}. Cu lines from the RF coil are also visible in these spectra.  The Knight shift from curves at 70 K and higher (including curves not shown for clarity of figure) combined with SQUID susceptibility~\cite{Boldrin2015} give an isotropic hyperfine coupling of $A^{\ce{^51V}}=\left( 0.64\pm0.03\right)\mathrm{\frac{T}{\mu_B}}$, which is somewhat lower than in Ba-vesignieite~\cite{Yoshida2012}. This result assumes that the $\ce{^{51}V}$ nucleus is coupled equally to all $\ce{Cu}$ atoms of a single hexagon.

\begin{figure}
  \includegraphics[width=0.9\linewidth]{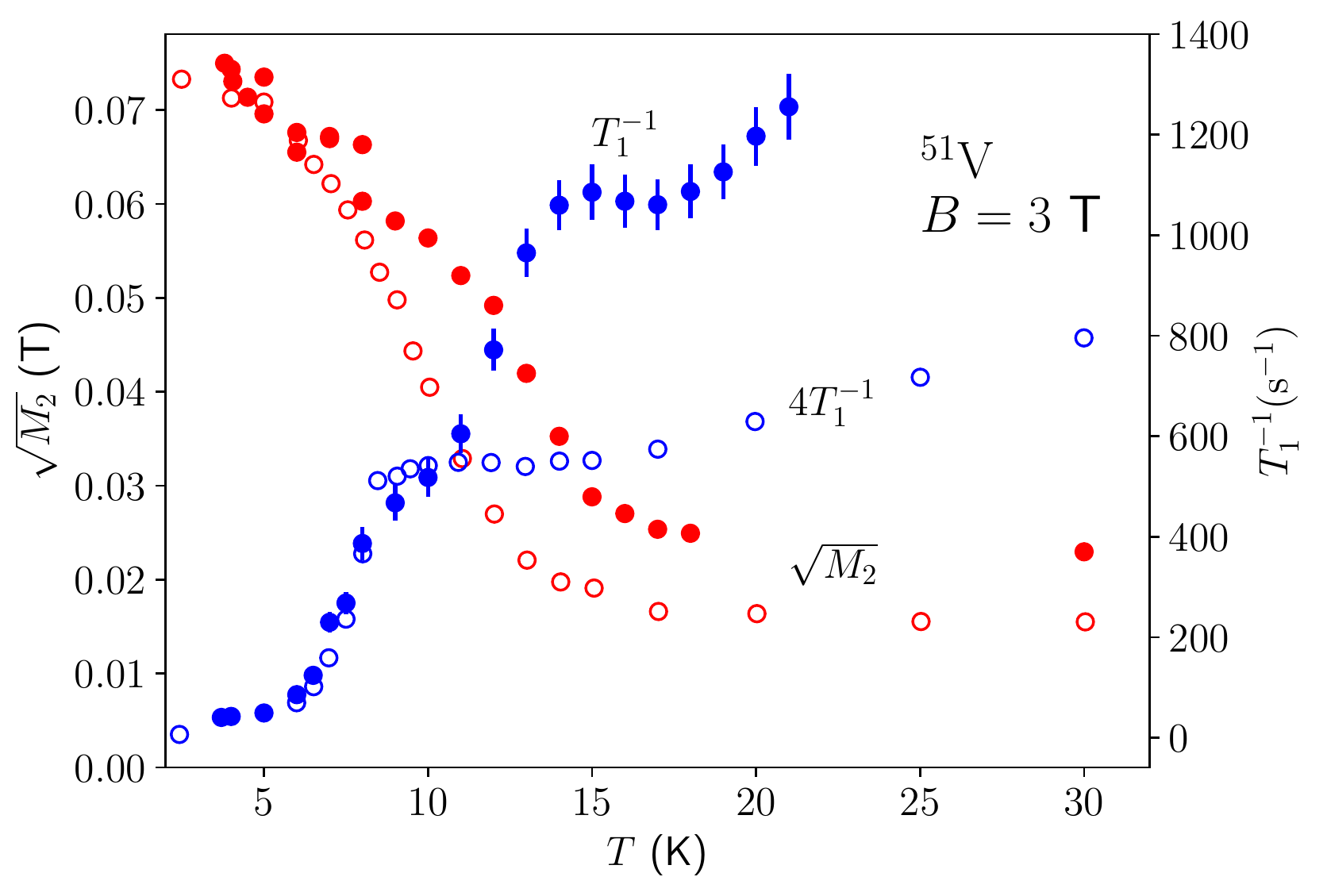}
  \caption{$\ce{^{51}V}$ NMR spectrum second moment (red) and $T_1^{-1}$ relaxation rate (blue) at varying temperatures. Sr-vesignieite data~\cite{temperature} (full circles) are compared with Ba-vesignieite data taken from~\cite{Yoshida2012} (empty circles).}
  \label{fig:T1_M2}
\end{figure}

As a function of temperature, the rate of broadening of the spectra can be seen to go through a maximum at around $13\pm 1$~K (see Fig.~\ref{fig:T1_M2}), which is close to the temperature at which a static component appears in $\mu$SR and an abrupt increase in susceptibility is observed~\cite{Boldrin2015}. As mentioned in the introduction, and illustrated in Fig.~\ref{fig:magnetic_orders}, a number of different ground states have been proposed for various KHAFM materials. In particular, Ba-vesignieite has at different times been proposed to have PVC~\cite{Yoshida2012}, NVC~\cite{Ishikawa2017} and hexagonal triple-$\vec{q}$~\cite{Boldrin2018} ground states. When the vector average of moments, $m_\mathrm{av}$, over a hexagon of Cu ions surrounding a $^{51}$V site is non-zero, one expects the usual powder spectrum of an ordered magnet: a square spectrum with width $\Delta f = m_\mathrm{av} 2\bar{\gamma} A \simeq m_\mathrm{av} \times 14$ MHz/$\mathrm{\mu_B}$. So, considering the amount of broadening observed, the vector sum of moments on a hexagon of Cu spins in the ordered phase must be nearly $\vec{0}$. Given the fact that the $^{51}$V nucleus is found in the center of a hexagon of Cu spins, this minimal amount of broadening allows us to narrow down the number of possible ground state configurations. In the case of coplanar SVC order, $m_\mathrm{av} = 0.5m$, and if we assume $m \simeq 0.5~\mathrm{\mu_B}$ or larger, this should give a width of at least 3.5 MHz, which is considerably larger than the main peak that is seen in Fig.~\ref{fig:triplePlots} \textbf{(c)}.

The case of hexagonal triple-$\vec{q}$ order is somewhat more complicated. Two possible chiralities should be considered, which we refer to as PVC-like and NVC-like (the NVC-like case is shown in Fig.~\ref{fig:magnetic_orders} and the PVC-like case in~\cite{Boldrin2018}). In both chiralities, the unit cell contains four hexagons. In three of these hexagons, the average moment is $m_\mathrm{av}=2m/3$ and in the fourth hexagon, $m_\mathrm{av} = 0$. The high $m_\mathrm{av}$ hexagons should give rise to a spectral width of 4.6 MHz or larger (for static moments that are at least $0.5~\mathrm{\mu_B}$ in magnitude), again much larger than the observed width. This makes it difficult to attribute a hexagonal triple-$\vec{q}$ state to Sr-vesignieite. At first glance, having V sites with different fields could explain what appears to be two peaks of different widths (3.7~MHz and 1.4~MHz) observable in Fig.~\ref{fig:triplePlots} \textbf{(c)}. However, this correspondence is unlikely since it would require very small moments of $0.4~\mathrm{\mu_B}$ and a 1:3 ratio (with larger weight for the larger width) of spectral weight between the two peaks, whereas in reality the spectral weight is closer to a 3:2 ratio.

In cases where $m_\mathrm{av}=0$, such as PVC and NVC, the width can be induced by dipolar fields or out-of-plane canting of the spins, as discussed in the following section. Thus, based on the small line width and assuming the static magnetic moment is not especially small (0.5~$\mathrm{\mu_B}$ or larger), SVC and hexagonal triple-$\vec{q}$ orders should be ruled out as potential ordering configurations of Sr-vesignieite and the two $\vec{q}=\vec{0}$ states are the most probable candidates.

Figure \ref{fig:T1_M2} displays the $\ce{^{51}V}$ NMR $T_1^{-1}$ relaxation rate and spectrum second moment as a function of temperature for the present sample and for Ba-vesignieite~\cite{Yoshida2012}. As is the case for Ba-vesignieite, an increase in the rate of broadening and a fairly abrupt drop in $1/T_1$ are observed in the vicinity of $T_\mathrm{N}$. However, there are notable differences between the behaviors of Ba-vesignieite and Sr-vesignieite. The first was expected from previous measurements: $T_\mathrm{N}$ is higher in Sr-vesignieite.  In Ba-vesignieite, the relaxation rate drops out roughly at $T_\mathrm{N}=9~\mathrm{K}$ but the broadening starts 4~K above $T_\mathrm{N}$. In Sr-vesignieite, both features appear at temperatures somewhat above the $T_N = 11$ K Néel temperature that was reported by Boldrin \emph{et al.}~\cite{Boldrin2015} based on the divergence of FC and ZFC susceptibilities. This suggests that the FC-ZFC splitting actually occurs well below the transition temperature in Sr-vesignieite. An additional hump in $1/T_1$ can be seen at around 9 K, but it is not clear that this is a real effect. It is more likely a statistical fluctuation or a slight anomaly in temperature control.

The sharp peak in $T_1^{-1}$ usually found in antiferromagnets is absent in both samples. This may be attributed to filtering of the $\vec{q}=\vec{0}$ mode. In the following expression for $1/T_1$~\cite{Moriya1963},
\begin{equation}
\left(T_1T\right)^{-1}=\frac{2\gamma_n^2\mathrm{k_B}}{g^2\mu_\mathrm{B}^2}\sum_{\vec{q}}\left| A(\vec{q})\right|^2\frac{\chi''_\perp\left(\vec{q},\omega_0\right)}{\omega_0},
\label{eq:Moriya}
\end{equation}
$\chi''_\perp\left(\vec{q},\omega_0\right)$ diverges at $T_\mathrm{N}$, which leads to the usual peak in the relaxation rate. This divergence may be compensated by $A(\vec{q})$ and the symmetry of the V site in a similar fashion to the reduced width of the spectrum: the magnetic field from the ordered Cu moments as well as its fluctuations are shielded at the V site by symmetry.

\subsection*{Canting Angle}

Of the initial magnetic orders that were shown in Fig.~\ref{fig:magnetic_orders}, only PVC and NVC seem to be consistent with the NMR results. However, the FC-ZFC splitting~\cite{Boldrin2015} in SQUID susceptibility indicates the presence of hysteresis in bulk magnetization, presumably originating from weak ferromagnetism. This section addresses the question of how canted versions of PVC and NVC can explain that phenomenon and how to estimate the magnitude of the canting angle from the NMR results. 

The canting that is most probable to occur in PVC magnetic order is out-of-plane canting. PVC should normally originate from a DM interaction forcing the KAFM to order and calculations indicate that PVC comes with out-of-plane canting when $D_z$ is positive and $D_p$ is non-zero or when $D_p$ is dominant over $D_z$~\cite{Elhajal2002}. One should keep in mind though that these calculations were made under the assumption of perfect kagome symmetry which constrains the $\vec{D}$-vector to being described by two components perpendicular to the Cu-Cu bonds, and this symmetry condition is not exactly true for Sr-vesignieite. This canting gives rise to weak ferromagnetism for the simple reason that the net moment in the unit cell is non-zero and perpendicular to the kagome plane, although it may be significantly smaller than the individual Cu moments. 

Assuming that all the broadening observed in the $\ce{^{51}V}$ spectra originates from isotropic hyperfine coupling and not from the dipolar interaction (this assumption will be justified later), we can find an expression relating the components of $\vec{m}$, the magnetic moment on Cu sites, to the hyperfine constants of V and Cu. The idea is to take the set of $\left(m_x,m_y,m_z\right)$ that satisfy a first constraint, the magnetic field on the Cu nucleus $B_{\mathrm{Cu}}$, and a second constraint, the magnetic field on the V nucleus $B_{\mathrm{V}}$. To simplify the notation, we will drop the superscript on the copper hyperfine interaction so that $A_{ij}^\mathrm{Cu} = A_{ij}$ from this point forward. The hyperfine interaction for the vanadium nucleus will be written as $A^V$ and assumed to be isotropic. Also note that we restrict our calculations to the on-site hyperfine interaction for Cu. 

Using $B_i=A_{i j}m_j$ with $i$ and $j$ representing space components:
\begin{equation}
B_{\mathrm{Cu}}^2=\sum_i\left(A_{i x}m_x+A_{i y}m_y+A_{i z}m_z\right)^2
\label{eq:Ellipsoid}
\end{equation}
If we choose a basis such that $\hat{z}$ is normal to the kagome plane and the principal axis of the Cu $d$-orbital is in the $xz$-plane, we can obtain the equation for an ellipsis by using $A_{x y}=A_{y x}=0=A_{y z}=A_{z y}$ and $m_z=\frac{B_\mathrm{V}}{A_\mathrm{V}}$:
\begin{align}
1=&\left(\frac{m_x-c_x}{E_x}\right)^2+\left(\frac{m_y}{E_y}\right)^2
\end{align}
Here the following definitions are employed:
\begin{align}
c_x=&-\frac{A_{x x}A_{x z}+A_{z x}A_{z z}}{A_{x x}^2+A_{z x}^2}\frac{B_\mathrm{V}}{A_\mathrm{V}}\\
E_x=&\sqrt{\frac{B_\mathrm{Cu}^2-\frac{B_\mathrm{V}^2}{A_\mathrm{V}^2}\left(A_{x z}^2+A_{z z}^2\right)}{A_{xx}^2+A_{z x}^2}-\left(\frac{c_x}{2}\right)^2}\\
E_y=&E_x\sqrt{\frac{A_{x x}^2+A_{z x}^2}{A_{y y}^2}}
\end{align}
These equations have a simple geometric representation. Since equation \ref{eq:Ellipsoid} describes an ellipsoid solution for $\vec{m}$ and $m_z=\frac{B_\mathrm{V}}{A_\mathrm{V}}$ describes a plane, their intersection describes an ellipsis. The components of $A_\mathrm{Cu}$ can be computed through a rotation of $63^\circ$ ($64^\circ$ in the case of Ba-vesignieite) around the $\hat{y}$ axis of the diagonal matrix $\left(\begin{smallmatrix}A_\perp & 0 & 0\\ 0 & A_\perp & 0\\ 0 & 0 & A_\parallel\end{smallmatrix}\right)$, since the shortest axes of the oxygen octahedra surrounding the Cu sites cross the kagome plane with angles of $27^\circ$. Using the $\ce{^{51}V}$ and $\ce{^{63,65}Cu}$ NMR results for $B_{\mathrm{Cu}}$ and $B_{\mathrm{V}}$ as well as calculated values for $A_\perp$ and $A_\parallel$ of $d_{x^2-y^2}$ and $d_{z^2}$ orbitals~\cite{Yoshida2007} allows us to define acceptable ranges for the canting angle (with respect to the kagome plane) and for the norm of the magnetic moment. These intervals arise from the degrees of freedom of the ellipsis solution and are not related to experimental uncertainties. The results of these calculations are given in Table~\ref{tab:angle_norm}.  Additionally, we have applied the constraint that the norm of the ordered magnetic moment be smaller than or equal to $1.19~\mathrm{\mu_B}$. This value is determined from $\left|\vec{\mu}_\mathrm{eff}\right|\simeq2.06~\mathrm{\mu_B}$ obtained in susceptibility measurements~\cite{Boldrin2015}. Since for a free spin-1/2 with projections $\vec{\mu}_\mathrm{eff}^z=\pm1\mathrm{\mu_B}$, $\left|\vec{\mu}_\mathrm{eff}\right|=g\mathrm{\mu_B}\sqrt{S(S+1)}$ becomes $\left|\vec{\mu}_\mathrm{eff}\right|\simeq\sqrt{3}\mathrm{\mu_B}$, a similar ratio would give $\vec{\mu}_\mathrm{eff}^z\simeq\pm1.19\mathrm{\mu_B}$. The constrained  moment sizes and canting angles are presented in parentheses in Table~\ref{tab:angle_norm}.

Table \ref{tab:angle_norm} shows that the intervals are much narrower in the scenario of a $d_{z^2}$ orbital, since the components of $A$ are much closer to each other in norm ($A_\perp=-12.2~\mathrm{\frac{T}{\mu_B}}$, $A_\parallel=10.2~\mathrm{\frac{T}{\mu_B}}$) than in the $d_{x^2-y^2}$ scenario ($A_\perp=2.71~\mathrm{\frac{T}{\mu_B}}$, $A_\parallel=-24.1~\mathrm{\frac{T}{\mu_B}}$). In all scenarios where the two compounds have the same unpaired orbital, the extremum values of moment norm are lower for Sr-vesignieite. This is due to the magnetic field at the Cu sites being of lesser magnitude in the Sr compound compared to the Ba one, which can be seen directly by comparing the zero-field spectra from this work with those of Yoshida \emph{et al.}~\cite{Yoshida2012}. In the case of $d_{z^2}$ orbitals, it is possible to conclude with confidence that the moments are smaller and the canting angle larger in Sr-vesiginiete. The smaller moment size could be explained by a reduced $D_z$ component of the DM interaction with respect to exchange interaction (thus reduced $D_z/J$ ratio), placing the material closer to the critical point between antiferromagnet and spin liquid~\cite{Cepas2008}. Meanwhile, a slightly increased $D_p/J$ in Sr-vesignieite may account for a slight increase in canting angle. For $d_{x^2-y^2}$ orbitals, this conclusion is not so robust as a change in the canting angle allows for significant changes in moment size.

\begin{table}
\begin{tabular}{ ccc cc c}
\hline
Compound & Orbital & \multicolumn{2}{c}{Canting angle ($^\circ$)} & \multicolumn{2}{c}{Moment ($\mu_{\mathrm{B}}$)} \\
 & & min & max & min & max\\
\hline
\multirow{2}{*}{Sr} & $d_{x^2-y^2}$ & 3.6 (5.4) & 33.3 & 0.20 & \quad 1.8 (1.19)\\
 & $d_{z^2}$ & 13.4 & 15.9 & 0.41 & 0.48\\
\hline
\multirow{2}{*}{Ba} & $d_{x^2-y^2}$ & 1.7 (4.0) & 14.6 & 0.32 & 2.8 (1.19)\\
 & $d_{z^2}$ & 6.4 & 7.5 & 0.63 & 0.74\\
\hline
\end{tabular}
\caption{Extremum values of canting angle and moment norm in the ellipsis solution for the case of PVC order.Results presented in parentheses are subject to the constraint that the moment norm not exceed $1.19~\mathrm{\mu_B}$. When the in-plane components of the magnetic moment point triangle to triangle (as illustrated in figure \ref{fig:magnetic_orders}), canting angle reaches a maximum in the $d_{x^2+y^2}$ scenario and a minimum in the $d_{z^2}$ scenario. Maximum values of the norm correspond to minimum values of the canting angle and vice versa.}
\label{tab:angle_norm}
\end{table}

\begin{figure}
  \includegraphics[width=0.9\linewidth]{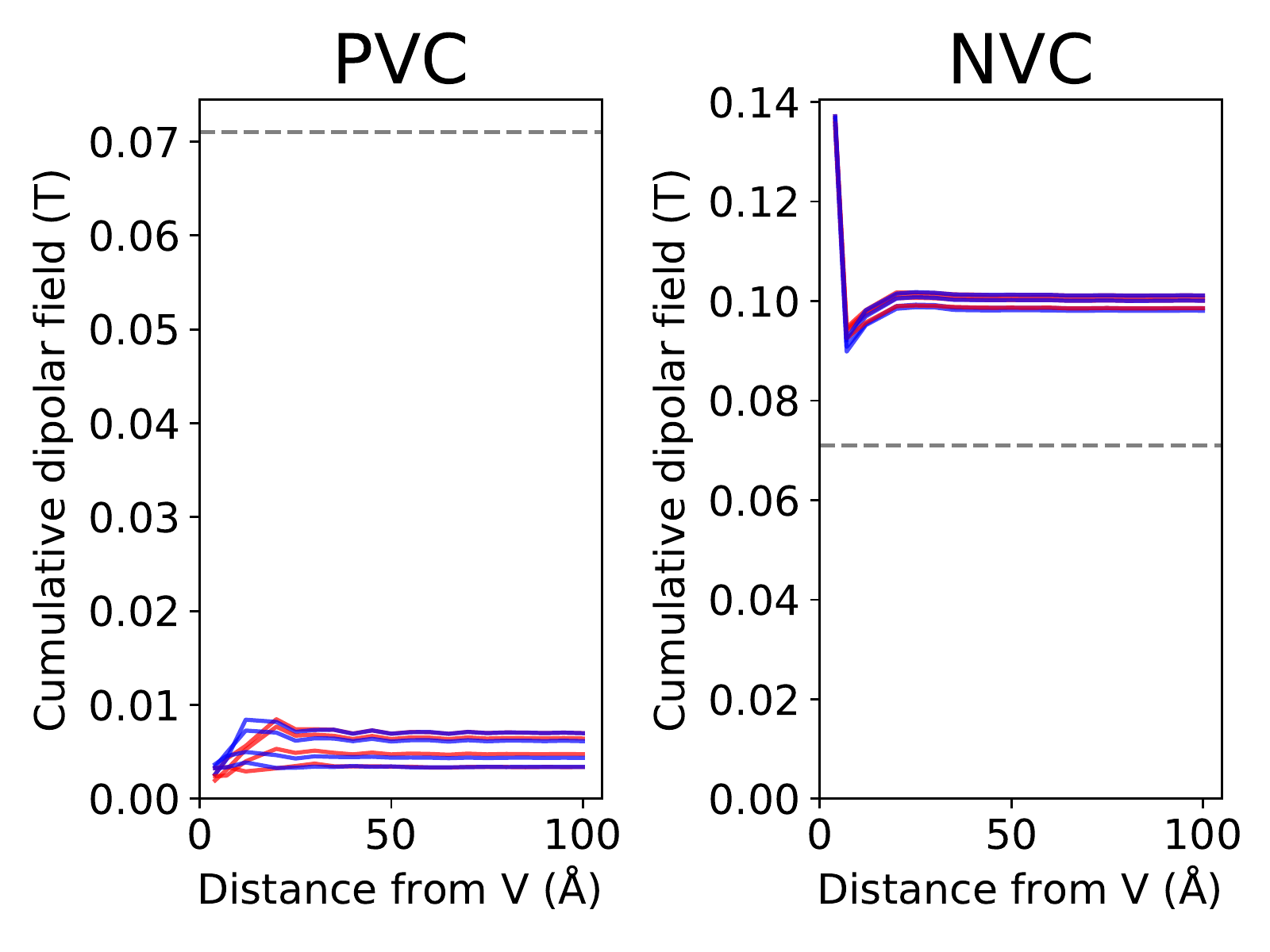}
  \caption{Simulation of the dipolar field at the two V sites (blue and red lines) due to  PVC and NVC magnetic orders in Sr-vesignieite with a finite number of unit cells, assuming point dipoles of 1~$\mathrm{\mu_B}$. Each line is calculated with a different global rotation of the order. A gray dashed line indicates the magnitude of the magnetic field observed experimentally on the V site. Only Cu sites located at a given distance or closer are included in the simulation. The cumulative field stabilizes at long distances when no canting is included. The magnetic orders are assumed to follow the $3_1$ screw axis symmetry.}
  \label{fig:comparative_dipolar}
\end{figure}

For NVC, in-plane canting has been observed in other systems~\cite{Okuma2017}. For a single kagome plane, in-plane canting leads to weak ferromagnetism in the same way as out-of-plane canting. However, in Sr-vesignieite, there is a $3_1$ screw axis symmetry~\cite{Boldrin2015}. If the magnetic structure respects this symmetry, it will lead to a 120$^\circ$ rotation of the total magnetic moment between planes, globally cancelling out the weak ferromagnetism of the individual kagome planes. Breaking of this symmetry due to interactions between the planes might nonetheless lead to a net in-plane canting.

Finally we turn to the role of the dipolar interaction in broadening the $^{51}$V spectra. Fig.~\ref{fig:comparative_dipolar} compares the magnetic field at the vanadium sites in PVC and NVC orders as a function of the distance over which it is summed. The latter induces a much higher magnetic field than the former. For $1~\mathrm{\mu_B}$ moments, the resulting dipolar field is 0.1~T. As opposed to the PVC scenario, the long-range dipolar field could account for a substantial part of the broadening of the $\ce{^{51}V}$ NMR signal. The canting angle could be considerably lower in the NVC scenario. This would not be surprising, as the in-plane canting in an NVC ground state results from a higher-order anisotropy term~\cite{Okuma2017}. Even without any weak ferromagnetism, $0.7~\mathrm{\mu_B}$ moments would give rise to an appropriate dipolar field at the vanadium site. With $d_{x^2-y^2}$ orbitals, this could be consistent with the measured zero-field Cu spectrum.  In the case of $d_{z^2}$ orbitals, the moment size must be smaller and some weak ferromagnetism would be required.

\section{Conclusion}

To summarize, we have confirmed with $\mu$SR and NMR the presence of long-range magnetic order below $T_\mathrm{N}$ as was suspected from thermodynamic measurements. These local magnetic probe techniques have provided additional information on the nature of that magnetic order, demonstrating that SVC and hexagonal triple-$\vec{q}$ structures are not compatible with the $^{51}$V NMR spectrum width. Furthermore, we suggest that the ordered moments are likely smaller in Sr-vesignieite than in Ba-vesignieite. This suggests that the Dzyaloshinskii-Moriya interaction is responsible for the ordering since its magnitude is understood to correlate with the size of the ordered moment~\cite{Cepas2008}. In that case, it furthermore implies that Sr-vesignieite is closer than its Ba counterpart to the critical value of $D/J$ that separates a moment-free phase (such as in herbertsmithite) from a Néel phase (such as in Ba-vesignieite)~\cite{Cepas2008}.

\begin{acknowledgements}
We are grateful to M. Lacerte and S. Fortier for technical support and A. Akbari-Sharbaf for fruitful discussions. We acknowledge research funding from the the Canadian grant agencies NSERC, CFI, FRQNT and CFREF, as well as Grant ANR-12-BS004-0021-01 ``SpinLiq''. 
\end{acknowledgements}


\begin{thebibliography}{29}
\bibitem[{tem()}]{temperature}
\bibinfo{note}{Some points at low temperature show two different values for the
  same $T$, hence these points should be considered with more uncertainty than
  the calculated ones that are smaller than the points. These uncertainties are
  suspected to originate from temperature control problems under 10~K.}
\end{thebibliography}
\end{document}